\documentstyle{l-aa}
\def\spose#1{\hbox to 0pt{#1\hss}}
\def\lta{\mathrel{\spose{\lower 3pt\hbox{$\mathchar"218$}}
     \raise 2.0pt\hbox{$\mathchar"13C$}}}
\def\gta{\mathrel{\spose{\lower 3pt\hbox{$\mathchar"218$}}
     \raise 2.0pt\hbox{$\mathchar"13E$}}}

\begin{document}  

\input epsf  

\thesaurus{ 05(05.03.01) ; 11(11.03.4)  }

\title{On two mass estimators for clusters of galaxies}

\author{H\'ector Aceves \& Jaime Perea }
\institute{Instituto de Astrof\'{\i}sica de Andaluc\'{\i}a, CSIC. Apdo. Postal 3004. Granada 18080, ESPA\~NA. }
\offprints{$\;$ H. Aceves $\;$ aceves@iaa.es}

\date{Received 20 July 1998 /  Accepted 4 February 1999 }

\maketitle

\begin{abstract}

	The two most common mass estimators that use velocity dispersions and positions of objects in a stellar system, the virial mass estimator (VME) and the projected mass estimator (PME), are revisited and tested using $N$-body experiments. We consider here only spherical, isolated, and isotropic velocity dispersion systems.

	We have found that the PME can overestimate masses by $\approx 20$\%, for realistic cluster mass profiles, if applied only to regions around the total system's {\it effective} radius. The VME can yield a correct mass at different radii provided that the potential energy term is correctly taken into account and the system is completely sampled; otherwise, it may lead to similar errors as the PME. 

	A surface pressure ($3PV$) term recently alluded to be usually neglected when using the VME and therefore required as a correction term is here found not necessarily required, although it can be used to yield a reasonable correction term. The preferred method here, however, is the virial theorem due to its simplicity and better agreement with $N$--body experiments. The possible reasons for the mass discrepancies found when using the PME and the VME in some $N$-body simulations are also briefly discussed.  


\end{abstract}

\section{Introduction}

Clusters of galaxies, probably the largest known virialized systems, provide an important tool to study the large-scale structure of the universe (e.g. Peebles 1980); for example, they provide an estimate for the dynamical mass density parameter on scales of $\sim$~1 Mpc. Clusters can also distort  the images of faint background galaxies by weak lensing providing a mean to study the cluster mass distribution and thus placing constraints on large-scale structure formation models (e.g. Miralda-Escud\'e 1995).

	Several methods are used to determine the mass of clusters of galaxies: e.g. velocity dispersions (Bahcall 1977); X-rays of the intracluster gas (Jones \& Forman 1984); and gravitational lensing (Tyson et al. 1990). In general, although  the three previous methods yield similar results for the total mass of rich clusters, $\sim 10^{14-15}{\rm M}_{\odot}$, they provide discordant values for the mass at different radii. This provides a motivation to have well characterized and tested methods for total mass and mass profile determination, owing to their importance e.g. in cosmology (White et al. 1993). We will deal here only with the first method mentioned.

	Analysis of the structural parameters in clusters of galaxies show that they are not really spherically symmetric structures, but  their shape appears to be somewhat triaxial (e.g. Plionis et al. 1991). Moreover, since some clusters still appear to be on the process of collapse, e.g. judging from the presence of substructures (West 1994), their velocity distribution is probably not isotropic. This may be particularly true at their outer parts where galaxies or material can be still infalling with predominantly radial orbits (White 1992).

	Nevertheless, observational limitations and simplicity makes us to assume virial equilibrium and sphericity for clusters when estimating their mass or mass profile. But even in such ideal scenario, methods that use observed radial velocities and projected positions are still a matter of some discussion. On the one hand, for example, Heisler et al. (1985, hereafter HTB) concluded that the projected mass estimator (PME) is probably the best option for estimating the mass of clusters of galaxies from galaxy motions. On the other hand, Perea et al. (1990) reached the conclusion that the best estimator is the virial mass estimator (VME) and that, for example, anisotropy, a mass-spectrum,  or the presence of substructure can lead to an overestimation of the mass if they are not properly taken into account.

	Furthermore, it seems that the above mass estimators do not even provide adequate masses when applied to $N$-body simulations; this is more critical, since in these systems we have complete information on the positions and velocities of particles.  For example, Thomas \& Couchman (1992, hereafter TC) using  $N$-body simulations of the formation of clusters of galaxies report that mass estimates based on the virial theorem underestimate the total mass by a factor $\approx 4$, and that the PME yields a factor $\approx 2$ too small. TC conclude that the PME is probably the best option to determine the mass using velocity dispersions.

	More recently Carlberg et al. (1997a, hereafter CYE) and Carlberg et al. (1996) have raised concerns regarding the consistency of the usual form of the VME when the total system is not entirely sampled. In their investigation of the average mass of galaxy clusters they establish that the VME overestimates their total mass by $\approx 20$\%, attributing this discrepancy to the neglect of the surface pressure ($3PV$) term in the continuous form of the virial theorem. A correction to virial mass estimates based on this $3PV$ term is starting to be applied to clusters of galaxies by the community (e.g. Girardi et al. 1998). Note that this finding of CYE is contrary to the trend observed by TC for the virial mass estimate.

	From our investigation of the VME and the PME presented below, we will show that although the $3PV$ term can account for the mass overestimation when  the VME is applied to a subsample of an equilibrium gravitational system, the correct application of the VME yields also a correct mass at different radii. The physical reason for the overestimation of mass by the VME in the previous situation lies in an incomplete consideration of the potential energy of the subsystem. We will show that when the potential energy is fully accounted for no discrepancy exists at any radius between the VME and the true mass for an $N$-body system. 
	In an astronomical situation, however, the previous result relies, of course, on having knowledge of the total extent of the system which is a somewhat ambiguous matter; this relates to the problem raised by CYE. 
	In practical terms, this also affects the use of the $3PV$ term as a correction factor since one needs a fair knowledge of the system's equilibrium extent, in addition to the velocity dispersion profile of a cluster.
	On other hand, the VME has the virtue of not depending on the form of the orbital distribution of galaxies in a near-equilibrium cluster, hence being straightforward to apply.

	Haller \& Melia (1996) have considered also the problem of calculating the mass profile by use of the PME. One of their conclusions is that a boundary term, arising from the finite sampling  of a gravitational system, may make appreciable contributions to the mass estimate at the inner regions of stellar systems. Their conclusion is similar to the one we reach here in an independent manner, but this boundary term is studied here to a larger extent using different `cluster' models and the physical reason for the overestimate of the mass is elucidated.

	In this work we revisit and test the PME and VME for an isolated spherical system of identical particles with an isotropic velocity distribution.
	We will quantify in particular the effect of a boundary term usually neglected in the PME and provide bounds to the errors that one may have when using the VME on a subset of an equilibrium system with different profiles. In a future paper we will apply our results to obtain mass profiles of nearby clusters and compare them with e.g. the profiles derived from X-rays.

	The plan of the paper is as follows. In $\S II$, the standard PME formalism is revisited and tested using different spherical models considered appropriate for the description of clusters of galaxies; e.g. those having a `cusp' and a `core'. A correction term is explicitly obtained and tested with $N$-body experiments.
	In $\S III$, the virial theorem, both in its discrete and continuous form, and the effect of the $3PV$ term on the mass determination are investigated. 	Finally, in $\S IV$, our main conclusions are presented.

\section{The Projected Mass Estimator}

	This estimator uses the average of the quantity $v_z^2 R$ over the whole tracer population (e.g. galaxies in a cluster) relating it to the total mass by (HTB, Perea et al. 1990)
\begin{equation}
M =\frac{f_\beta}{G} \frac{1}{N} \sum_i v_{z_i}^2 R_i =  \frac{4-2\beta}{4-3\beta} \frac{32}{\pi G} \langle v_z^2 R \rangle  \;,
\end{equation}
where $v_{z_i}$ is the observed line-of-sight velocity of a galaxy relative to the cluster mean, and $R_i$ is its projected radius from the centre of the distribution. The factor $f_{\beta}$ is related to the orbital distribution of galaxies and therefore a function of the anisotropy parameter $\beta$ (see below), being equal respectively to $64/\pi$ and $32/\pi$ for the cases of radial ($\beta=1$) and isotropic ($\beta=0$) orbits.

	Equation (1) is obtained as follows. On the one hand, for a spherically symmetric system under steady-state conditions and no rotation the Jeans equation holds (Binney \& Tremaine 1987)
\begin{equation}
\frac{{\rm d}}{{\rm d}r} \rho \sigma_r^2 + \frac{2\beta}{r} \rho \sigma_r^2= -\rho \frac{{\rm d}\varphi}{{\rm d}r}   \;,
\end{equation} 
where $\rho$ is the mass density, $\varphi$ the gravitational potential,  $\sigma_r$ the radial velocity dispersion, and $\beta=1-\sigma_t^2/\sigma_r^2$ the anisotropy function; $\sigma_t$ is the tangential velocity dispersion. Multiplying Jeans equation by $r^4$, and integrating by parts up to a radius $r$ we have:
\begin{equation}
r^4 \rho \sigma_r^2 -
\int_0^{r} r^3 \rho \sigma_r^2 (4 - 2\beta) \,{\rm d}r
 = -\int_0^{r} r^4 \rho \frac{{\rm d}\varphi}{{\rm d}r} \,{\rm d}r    \;. 
\end{equation}

	On the other hand, the average $\langle v^2_z R\rangle$ in Eq. (1) up to a particular radius $\,r$ is estimated as (e.g. HTB):
\begin{equation}
\langle v^2_z R\rangle_r = \frac{ \int^r v_z^2 R \, f({\mathbf r}, {\mathbf v}) \, {\rm d}^3{\mathbf r}\, {\rm d}^3{\mathbf v} }{M(r)} 
= \frac{ \int^{r} \rho
 \sigma_z^2 R \,{\rm d}^3 \mathbf{r} }{M(r)}  \;,
\end{equation} 
where $\sigma_z^2 = \sigma_r^2 \cos^2 \theta + \sigma_t^2 \sin^2 \theta$ is the velocity dispersion along the line-of-sight, $R= r \sin \theta$, and  $f({\mathbf r}, {\mathbf v})$ is the phase space density, defined such that 
$f({\mathbf r}, {\mathbf v})  {\rm d}^3{\mathbf r} {\rm d}^3{\mathbf v}$ is the mass contained in the phase space volume element ${\rm d}^3{\mathbf r} {\rm d}^3{\mathbf v}$.

	Evaluating the integral in Eq. (4), by using Eq. (3) and \emph{assuming isotropy} ($\beta=0$), we have:
\begin{equation}
\langle v^2_z R\rangle_r =
\frac{\pi^2}{4M(r)} \left[\,r^4\rho(r)\sigma_r^2(r)
 + \int_0^r r^4 \rho(r)\frac{{\rm d}\varphi}{{\rm d}r} {\rm d}r  
 \, \right]  .
\end{equation} 
Since ${\rm d}\varphi = G M(r) {\rm d}r/r^2$ and $4 \pi r^2 \rho = {\rm d}M/{\rm d}r $ we have, after rearranging terms:
\begin{equation}
\Delta (r)\equiv \frac{32}{\pi G}\langle v^2_z R\rangle_r - M(r) = \frac{8\pi}{G M(r)} \,r^4\rho(r) \sigma_r^2(r) \;.
\end{equation}
This function $\Delta(r)$ gives us directly the effect of neglecting the boundary term on the estimation of the mass via the PME when partial sampling is done. Now, since one usually fits a particular projected model to a set of data, the quantities to evaluate $\Delta (r)$ follow from the corresponding deprojected model. One may use e.g. Eq. (9) to evaluate the radial velocity dispersion in case this does not exists explicitly beforehand for the adopted model. Integrating over the whole system we recover the standard formula of the projected mass for an isotropic stellar system: $M_{\rm P}=32\langle v^2_zR\rangle /( \pi G )$.

	Therefore, in using equation (1) for an isotropic system one must remember that this form of $M_{\rm P}$ holds only when the total extent of the stellar system has been sampled; otherwise, the correction term $\Delta (r)$ has to be applied.
	As will be shown below, in astronomical situations the use of $M_{\rm P}$ can yield a significant overestimate of the system's mass especially if we sample only around its {\it effective} radius.
	Furthermore, since the boundary of an astronomical system is somewhat ambiguous, particularly if large amounts of dark matter exist at the outer optical edges, it is important to quantify the error one might commit  when using the PME. We proceed to estimate such error in the next subsection.

\subsection{Boundary Term for Spherical Systems}

In this subsection we estimate theoretically the boundary term $\Delta (r)$ for four different spherical models: (1) de Vaucouleurs' or $R^{1/4}$ (de Vaucouleurs 1948), (2) Hernquist's (1990), (3) Dehnen's ($\gamma=0$) (1993), and (4) a King's modified profile. We consider these models to bracket most of the possible profiles that can be fitted to a cluster of galaxies, which will be here the astronomical object of interest. Evidently, these profiles can be applied to other spherical systems, as well as the mass estimators.

Navarro, Frenk \& White (1996), hereafter NFW, have proposed a profile model appropriate for clusters of galaxies of the form $\rho(r) \propto r^{-1} (r+a)^{-2}$, with $a$ being a scale radius. Carlberg et al. (1997b) have shown that the average mass density profile of clusters is well described by the  NFW profile though they mention that the Hernquist's model is statistically acceptable. However discordant results exist in the literature on the subject of cluster profiles (e.g. Adami et al. 1998). We have not used the NFW profile here due to its intrinsic difficulty in defining formally a total mass, hence a half-mass or effective radius, unlike the other profiles.  Indeed, in the NFW profile $\rho \to r^{-3}$ as $r\to \infty$ and the total mass diverges. We do not expect much difference in the boundary term $\Delta (r)$ when the NFW  and Hernquist profiles are fitted to real clusters. Thus, for convenience, we deal here only with the Hernquist's model.

Here the units adopted for the numerical calculations are such that the total mass is unity, $M=1$, the total system's {\it effective} radius $R_{\rm e}=1$, and the gravitational constant  $G=1$. Following, for completeness, we write some pertinent quantities to compute $\Delta (r)$ for the different models to be considered.

\subsubsection{De Vaucouleurs Model}

The surface brightness of the de Vaucouleurs' model is:
\begin{equation}
I(R)=I_{\rm e} \exp\{\,-b\,[\,(R/R_{\rm e})^{1/4} - 1]\,\} \;,
\end{equation}
where $R_{\rm e}$ is the {\it effective} radius, a scale-radius where half of the total light is emitted, and $b=7.66925$. We will consider here the effective radius to be the locus where half of the projected mass resides.
The total mass is, assuming a mass-to-light ratio of unity, $M = 8! \pi \exp(b) R_{\rm e}^2 I_{\rm e}/b^8\,\approx 22.7 R_{\rm e}^2 I_{\rm e}$.

The mass density $\rho(r)$ can be obtained by deprojecting the surface brightness using the Abell integral equation (Binney \& Tremaine 1987).
The latter has been obtained in implicit form by Poveda et al. (1960) as:
$$
\rho (r) = \frac{2 I_{\rm e} b \exp(b)}{\pi}\, j^{-3} \, \int_1^{\infty} \, 
{{{\exp(-b jt) \sqrt{t-1} } \over {\sqrt{(t+1)(t^2+1)  (t^4+1)}}}}\, 
$$
\begin{equation}
\times \Bigl\{ b j \,+\, {1 \over {2(t+1)}} + {t \over {(t^2+1)}} +{2t^3\over{(t^4+1)}}\Bigr\}\,{\rm d}t  \,,
\end{equation}
where $t=R^{1/4}$, and $j=r^{1/4}$.

The product $\rho \sigma_r^2$ can be obtained from the Jeans equation by imposing the condition that $\rho \sigma_r^2 \to 0$ as $r\to \infty$. Since $\rho$ and ${\rm d}\varphi/{\rm d}r$ are non-negative, and $\sigma_r^2$ always positive, we have:
\begin{equation}
\frac{{\rm d}\,\rho \sigma_r^2}{{\rm d}r}=-\rho \frac{{\rm d}\varphi}{{\rm d}r}
\;\;\to\;\; \rho(r)\sigma_r^2(r)=\int_r^{\infty} \frac{M(r)\rho(r)}{r^2}\,{\rm d}r \,.
\end{equation}
The mass $M(r)$ follows from integrating the mass density, and the boundary term can then be computed numerically.

\subsubsection{Hernquist Model}

This model is similar to an $R^{1/4}$ model, but has the virtue of being expressed in terms of simple functions. Some of the quantities of interest here are:
\begin{equation}
\rho(r)=\frac{M a}{2 \pi} \frac{1}{r(r+a)^3} \;,\quad 
M(r)= M \frac{r^2}{(r+a)^2} \;,
\end{equation}
where $a$ is a scale-radius, and $M$ the total mass. The effective radius here is $R_{\rm e}=1.8153a\,$.

Using Hernquist's expressions and integrating several times by parts equation (9), we find that:
\begin{equation}
\frac{\rho(r) \sigma_r^2(r)}{G M^2 a /2\pi } = - \left\{ \sum_{\lambda=1}^4
 \frac{a^{\lambda - 5}}{\lambda (r+a)^{\lambda}}+ \frac{1}{a^5} \ln \frac{r}{r+a}\right\} \,.
\end{equation}

\subsubsection{Dehnen $\gamma=0$ Model}

Dehnen (1993) has provided a family of potential-density models for spherical systems. We use his $\gamma=0$ model since it provides a core resembling King models and allows simple analytic calculations (see also Tremaine et al. 1994, model $\eta=3$). The density and mass are given respectively by:
\begin{equation}
\rho(r) = \frac{3Ma}{4 \pi} \frac{1}{(r+a)^4} \;,\quad
M(r) = M \frac{r^3}{(r+a)^3} \,.
\end{equation}
The velocity dispersion is 
$\sigma_r^2 = G M (a+6r)(r+a)^{-2}/30$.\footnote{A misprint in Dehnen's Eq. A3 is corrected here.} The effective radius in this model is $R_{\rm e}=2.9036a\,$.

\subsubsection{King Modified Model}

The simplicity of this empirical model, besides allowing larger cores, makes it well suited for fitting observational data. The projected and physical density, and projected mass are (e.g. Perea et al. 1990, Adami et al. 1998), respectively:
\begin{eqnarray}
I(R) & = & I_0  \left[ 1 + \left( \frac{R}{R_0} \right)^2 \right]^{-\alpha} , \\
\rho(r) & = & \rho_0 \left[ 1 + \left( \frac{r}{R_0} \right)^2 \right]^{-(\alpha+1/2)} \nonumber
\end{eqnarray}
\begin{equation}
M(R) =  M\, \left\{\;
1 - \left[ 1 + \left( \frac{R}{R_0} \right)^2  \right]^{(1-\alpha)} \;\right\},
\end{equation}
where $\alpha$, and $R_0$, the {\it structural length} of the model, are fitting parameters. The effective radius, in terms of $R_0$, is $R_{\rm e}^2 = R_0^2\, [ 2^{1/(\alpha -1)} - 1 ]$. The total mass $M$ may be estimated e.g. by the virial theorem or considered another fitting parameter. The following relations among the different parameters hold: 
$$
M = \frac{1}{\alpha -1}\,\pi I_0 R_0^2 \,, \quad
\rho_0 = \frac{\Gamma(\alpha + 1/2) }{ \sqrt{\pi}\,\Gamma(\alpha)} \frac{I_0}{R_0} \,, 
$$
$\Gamma$ being the Gamma function. Hereafter we will consider only the value of $\alpha=1.3$, which is close to fittings to the Coma cluster profile (e.g. Perea et al. 1990) and to the spectroscopic value of a $\beta$--model of X--ray emission in clusters (Sarazin 1988). We refer hereafter to this model as King.

\subsubsection{Theoretical Value of the Boundary Term}

In Fig. 1 we show the function $\Delta (r)$ for the above considered models. 
Of the models considered here the Dehnen's  model lends itself to an easier computation of the maximum value of the boundary term. Indeed, the following expression for the boundary term is readily obtained
\begin{equation}
\Delta (r) = \frac{G M a \,r\, ( a + 6 r)}{5 \,(r + a)^3} \,.
\end{equation} 
The maximum of $\Delta(r)$ occurs at $r_{\rm max}=(5+\sqrt{31})a/6$, or $\approx 0.61$ for our chosen units, and has a value of $0.194\,$.
For the de Vaucouleurs' model the maximum value of this boundary term is $0.166$,  for the Hernquist's one $0.176$, and for the King's $0.21\,$. The radius at which these three maxima occur are at 1.07, 0.66,  and 0.43, respectively; i.e. all occur at  $r\approx R_{\rm e}\,$.

	A trend for the maximum of $\Delta (r)$ to increase as the system gets less concentrated is observed. The latter is explained if we consider the extreme case of an isothermal sphere, where $\rho \propto \sigma_r^2 r^{-2}$, $M \propto \sigma_r^2 r$, and $\sigma_r^2$ is constant, from where we obtain that $\Delta (r) \propto r\,$. Therefore  $\Delta_{\rm max}$ grows as the `isothermal' region of the model grows. 
	Since the boundary term is positive and linear in $M$ we can be overestimating the total mass by about 20\% by using the quantity $32 \langle v^2_zR\rangle/(\pi G)$ as in Eq. (1, $\beta=0$), if the sampling is done only up to regions near the effective radius of the gravitational system. The precise value of the error depends on the density profile of the system.

\begin{figure}[!t]
\epsfysize=6.5cm
\epsfxsize=8.5cm
\epsffile{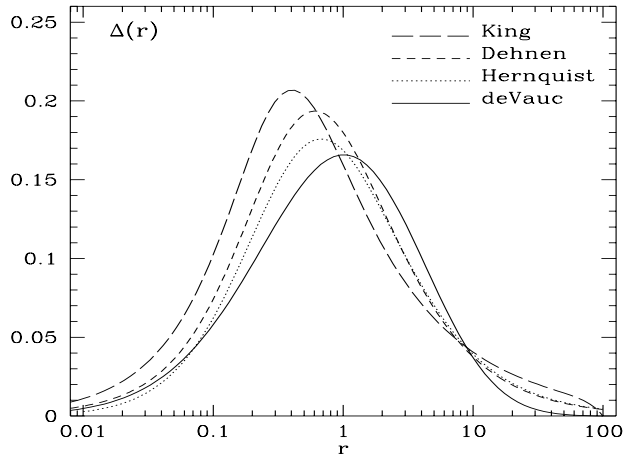}
\caption{Boundary term in the PME as a function of radius for de Vaucouleurs', Hernquist's, Dehnen's, and King's models. The maximum values occur respectively at $\approx (0.17,0.18,0.19,0.21)$ and inside $R_{\rm e}\,$. These values give an indication of the error that one may expect to make when estimating masses via the PME at different radii of e.g. a cluster of galaxies.}
\end{figure}

\subsection{Numerical Test of the PME}

	Here we evaluate $M_{\rm P}=32\langle v^2_zR\rangle/( \pi G )$ as a mass estimator by using $N$-body models. We only present here for convenience the results for the Dehnen's model, but the same quantities and analogous results consistent with the theoretical expectations were found for the other three models of $\S2.1$.

	A Monte Carlo realization of $N=10^4$ particles, each of mass $m_i=1/N$, was constructed. The numerical `cluster' was first evolved using a TreeCode (Barnes \& Hut 1986) in order to test its stability before the PME is applied to the initial configuration.  The total time of evolution lasted 160 time-steps, each of $\Delta t =0.031$ units, with  softening parameter $\epsilon =0.025$ and tolerance parameter $\theta = 1$. Quadrupole terms were included. Energy changed by less than 0.1\% throughout the evolution.

	The ratio of kinetic to potential energy, as provided by the code, was initially $T/U = -0.52$ reaching  $T/U=-0.50$ at the end of the computation.  The initial discrepancy of the numerical model from the ideal virial ratio is attributed to the initial positions and velocities not being in equilibrium with the code's potential which depends on the softening introduced: i.e. $\varphi \propto {(r+\epsilon)^{-1}}$. After this test was conducted, we considered our initial $N$-body system to be in stable equilibrium so we could apply to it the mass estimators considered in this work.

	To compute $32\langle v^2_z R\rangle/( \pi G )$ we proceeded as one would do observationally to compute the mass at different radii of clusters. We divided the $N$-body cluster into 40 projected concentric radii, spaced logarithmically, from $(0.1 \to 100) R_{\rm e}\,$. Inside each radius the summation of $v_{z_i}^2 R_i$, with $R_i = (x_i^2+y_i^2)^{1/2}$, was calculated for all pertinent particles and then divided by their number $N(R_i)$. In this way we estimated $32\langle v^2_zR\rangle/( \pi G )$  or, equivalently, the observational projected mass $M_{\rm P}$.

	In Fig. 2 we plot $M_{\rm P}$ as a function of projected radius for our Dehnen numerical cluster. Also shown are: the virial mass ($M_{\rm V}$), as discussed below, the true projected mass $M(R)$, and the real 3-D mass within radius $r$, $M(r)$. In the upper panel of Fig. 2, the difference $M_{\rm P} - M(r)$ is shown as a long-dashed line, the difference $M_{\rm P}-M(R)$ as a dotted line, and the theoretical value of $\Delta (r)$ as in equation (15) is shown as a solid line.	The difference between $M_{\rm P}$ and $M(r)$ can be said to be in perfect agreement, both in magnitude and in position of the maximum, with the theoretical expectations of $\S 2.1$. Poisson noise can explain the obtained differences as we have verified, although it is not shown in the figure to avoid overcrowding.	The difference $M_{\rm P} - M(R)$, although similar in shape to $\Delta (r)$, is about 50\% smaller in magnitude. A similar behaviour of these discrepancies was observed for the other models considered here.

\begin{figure}[!t]
\epsfysize=8.5cm
\epsfxsize=8.5cm
\epsffile{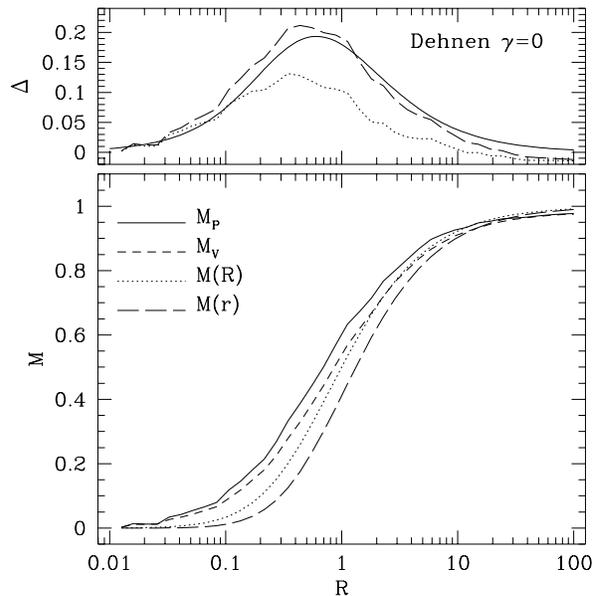}
\caption{Results of the application of the PME ($M_{\rm P}$) and VME ($M_{\rm V}$) as a function of radius to Dehnen's ($\gamma=0$) numerical model of a cluster. Also, the theoretical projected mass profile $M(R)$ and physical mass $M(r)$ are shown for comparison. The upper panel shows the differences $M_{\rm P} - M(r)$ ({\it long-dashed}), $M_{\rm P} - M(R)$ ({\it dotted}), and the theoretical boundary term $\Delta (r)$ is plotted as a solid line. } 	
\end{figure}

\subsection{Intrinsic `Error' in the PME}

	From the analysis of the boundary term in Eqs. (6,15) and the numerical calculations performed, we conclude that the PME can overestimates the mass by $\approx 20$\% of its true value if the sampling is made only to $\approx 1 R_{\rm e}$ of the gravitational system. This maximum error is for the King's model, while lower values were obtained for more concentrated models like Hernquist's or de Vaucouleurs' in agreement with the theoretical expectations; see Fig. 1.

	It is important to recall at this point the assumptions made in the derivation of the PME: spherical symmetry, isotropic velocity distribution and no mass spectrum; implicitly it also assumes that the particles we {\it observe} trace matter.  If dark matter dominates the outer parts of galaxy clusters we expect that sampling the luminous part  would yield an overestimate in the mass, not an underestimate,  which can be of $\approx 20$\%. This particular value depends, obviously, on the total mass profile, but can be considered typical for profiles suited for galaxy clusters.

	Before we leave this section we address the results of Thomas \& Couchman (1992) on the PME. Examining their formulae $B2$ we note that these equations correspond to the case of test particles orbiting a massive particle, as they acknowledge, but for non-test particles they would require an extra factor of two. 	It seems, from the numbers they give in their Table 5 that this extra factor of two and perhaps a higher constant anisotropy, e.g. $\beta = 0.75$, would help solving the discrepancy they mention,  although it is not clear that one can characterize by a constant anisotropy their $N$-body clusters.

\section{The Virial Mass Estimator}

In this section we test the virial theorem as a mass estimator. We consider its discrete and continuous form in order to understand and clarify some matters regarding its use that have recently appeared in the literature (e.g. Carlberg et al. 1996). We first outline the deduction of the scalar virial theorem for the sake of completeness and future reference (see  e.g. HTB).

\subsection{Discrete Virial Theorem}

Consider a system of $N$ equal-mass particles interacting gravitationally. First, we differentiate twice the moment of inertia of the system $I= (m/2) \sum {\mathbf r}_i \cdot {\mathbf r}_i$ with respect to time:
\begin{equation}
\frac{{\rm d}^2 I}{{\rm d} t^2} = m \sum_i \frac{{\rm d}^2 {\mathbf r}_i}{{\rm d}t^2} \cdot {\mathbf r}_i 
+ m \sum_i \left( \frac{{\rm d} {\mathbf r}_i }{{\rm d}t} \right)^2 \,.
\end{equation}
The second term on the right-hand side is twice the total kinetic energy of the system, and the first term is the potential energy
\[
m \sum_i \frac{{\rm d}^2 {\mathbf r}_i}{{\rm d}t^2} \cdot {\mathbf r}_i 
= - G m^2 \sum_i \sum_{i<j} \frac{1}{r_{ij}} \,,
\]
where $r_{ij}=|{\mathbf r}_i - {\mathbf r}_j|$.

	To derive the common form of the virial theorem we require that the time average of the second derivative of the moment of inertia vanishes (e.g. Goldstein 1980).  In this case we have, after averaging over all possible orientations and assuming an isotropic velocity distribution (Limber \& Mathews 1960), that 
\begin{equation}
M_{\rm V} = \frac{3 \pi N}{2 G} \frac{\sum_i v_{z_i}^2}{\sum_{i<j} 1/R_{ij}} \,,
\end{equation}
where $R_{ij}$ is the projected separation among pairs of particles. This is what we explicitly mean by the virial mass estimator, VME. 
	An important observational advantage of the VME in determining the total mass of a system is that it has virtually no dependence on velocity anisotropy  for near-spherical systems (e.g. The \& White 1986).

	One may also derive a virial estimator by using the distances of test particles to the centre of the  system (Bahcall \& Tremaine 1981). Nonetheless, such form of the virial theorem is of no interest here since we do not consider galaxies in clusters to be properly represented by test particles.

\subsubsection{Numerical Test of the VME}

	We applied the VME to the numerical models described in $\S2$. 
We proceeded as in the testing of the PME using different aperture radii, where  we applied the VME to the particles inside the corresponding aperture. The mass derived using the VME, $M_{\rm V}$, is shown also in Fig. 2 for the Dehnen's model. Note that the VME follows more closely the projected mass $M(R)$ than the PME over most radii, especially for $R \gta R_e\,$.

	Nonetheless, significant deviations occur at $R \lta R_{\rm e}\,$. In Fig. 3  we show for the Dehnen's model the difference between the  VME estimate and the physical mass at different radii ({\it short-dashed}), $\delta_{\rm V}$, and that of the non-projected virial estimator with respect to $M(r)$ ({\it solid}), $\delta_{\rm 3V}$.  The maximum differences are  $\delta_{\rm V} \approx 15$\% and $\delta_{\rm 3V} \approx 12$\%. For the King profile we obtained a somewhat higher $\delta_{3\rm V} \approx 20$\%, this since it has a shallower profile than the Dehnen's one.
	For the Hernquist's model we obtained lower values in general: $\delta_{\rm V} \approx 12$\% and $\delta_{\rm 3V} \approx 10$\%. Also shown in Fig. 3 is the  boundary term (15) that appeared in the PME ({\it dot-dashed}).

\begin{figure}[!t]
\epsfysize=6.5cm
\epsfxsize=8.3cm
\epsffile{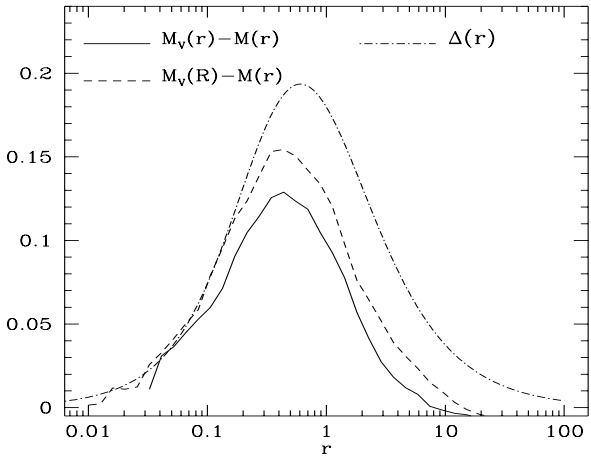}
\caption{Differences in virial mass at different radii, both projected and non-projected, with respect to the true mass of the Dehnen's model are shown. Also, the boundary term $\Delta (r)$ from the PME is shown.}
\end{figure}

	The application of the VME to our simulated $N$-body system does not yield any underestimate of the true mass. On the contrary, overestimates for the mass at the inner regions of the numerical systems were found.  These results are contrary to those found by  TC in their application of the VME, but in the same direction as noted by CYE.  We will now turn to explain the reason why we have such discrepancies in the result from our recent application of the VME and $M(R)$ at different radii.

	Here we recall that the basic idea behind the use of the virial theorem as a mass estimator is to relate the gravitational potential energy and the kinetic energy of the system. If the complete system is in virial equilibrium any subsystem has to be also in equilibrium.  Computing of the kinetic energy poses no problem, but the gravitational energy, i.e. the term involving $\sum 1/R_{ij}$, has not been correctly considered earlier when we computed $M_{\rm V}(R)$.

	Unlike force, the potential energy at a particular point $\tilde{r}$ inside a sphere depends on the particles outside this radius. 
	In other words, the virial theorem holds at all radii if the \emph{total} potential energy of the particles inside this  $\tilde{r}$ is accounted for. As said before, the kinetic part of the virial theorem does not suffers from this inconvenient. All this is important to take into account when mass profiles of clusters are to be computed and compared e.g. to mass profiles derived from $X$-ray observations or gravitational lensing.

	Therefore, applying the VME to clusters, for which we probably do not have a fair value of its true boundary, can yield an overestimate of the mass because one ignores the contribution to the potential energy of the outer parts of the `total' cluster which may be dark. 
	The neglect of this extra potential energy is transformed apparently as a mass excess inside the sampled cluster, since the system requires this extra mass to be in equilibrium in order to equate the corresponding kinetic energy. 
	This is also equivalent to having particles of different masses as we increase the aperture in the system (heavier particles inside and lighter ones outside). 	We need to consider a surface pressure term in the VME (17) if e.g. the system were confined by an external force, but this is not the case here. Note that we do not take into account here the probable effect of external tidal fields on the mass determination of clusters of galaxies.

\begin{figure}[!t]
\epsfysize=8cm
\epsfxsize=8.5cm
\epsffile{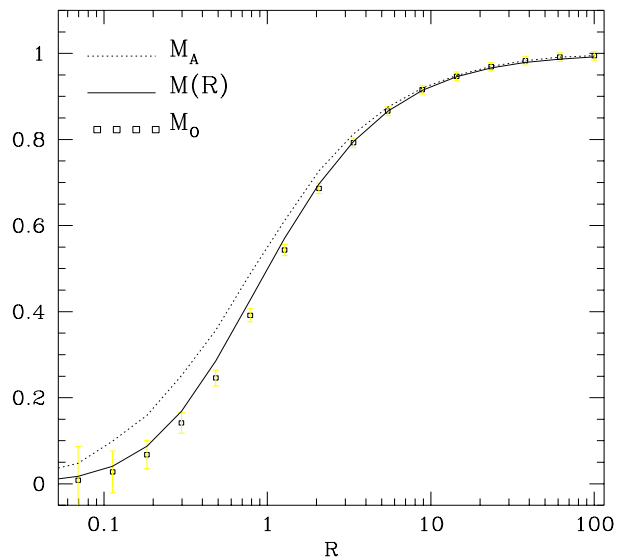}
\caption{Projected mass profile derived from the virial theorem (eq. 17) applied to an $N$-body equilibrium and isotropic Dehnen system. $M_{\rm A}$ assumes that the summation of interparticle distances is constrained to particles inside a particular radius $\tilde{R}$. The solid line indicates the corresponding projected mass, and  $M_{\rm O}$ considers correctly that the particles outside  $\tilde{R}$ contribute also to the potential energy of particles inside $\tilde{R}$.}
\end{figure}

	In Fig. 4 we show the results of a consistent calculation of the potential energy in the VME (eq. 17) that supports our previous qualitative explanation.  We consider here only the projected form of this virial estimator owed to its use in observational astronomy; an almost perfect agreement was found when we used its non-projected form.  In dotted line we show the mass estimated using the virial theorem by considering particles only within a certain `aperture' radius $\tilde{R}$; i.e.  the summation of inter-particle separations in Eq. (17), for each $i$-th particle, considers only the $j$-th particles that their physical position is $R_j \le \tilde{R}$.  The projected mass of the Dehnen's model is shown as a solid line, and the mass estimated using what the theory of the virial theorem indicates, $M_{\rm O}$, is shown in open squares; i.e. the potential energy of the $i$-th particle inside this $\tilde{R}$ is computed by adding the contributions of all other particles to $\sum 1/R_{ij}$, including those that have $R_j \ge \tilde{R}$.

	As observed from Fig. 4, the two above procedures yield  different results, except for small discrepancies which we attribute to numerical artifacts in the construction of the $N$--body system. The error bars in Fig. 4 correspond to Poisson fluctuations $\propto 1/\sqrt{N}$, and the maximum discrepancy is $\approx 0.04$ which consistent with the numerical noise.
	Similar differences were observed for the other models considered here. For systems with less particles the behaviour is the same, but of course the errors are larger.

	The physical reason for the overestimation of the mass in the application of the PME is also now clear: it does not account for the long-range nature of gravity. The product $v_{z_i}^2 R_i$ in the PME only considers the distance of a galaxy to the centre of the mass  distribution and does not account for its interaction with other particles, as the summation involving $1/R_{ij}$ in the VME does.

	The behaviour shown in Fig. 4 supports the idea that the VME is an excellent mass estimator.  In real applications,  we can expect the error due only to the VME method to be within $\approx 20$\% for realistic mass profiles if sampling is made  around the total effective radius of the cluster.  
	On other hand, taking $M_{\rm V}$ ($M_{\rm O}$ in Fig. 4) as a good estimate of the mass profile its difference with $M_{\rm P}$ may indicate us how far into the equilibrium part of a gravitational system are we sampling, and thus lower the previous error bound. It might also tell us something about its velocity anisotropy.

	In astronomical applications mass estimates are somewhat unreliable due to the fact that clusters may not be in virial equilibrium and/or due to the presence of interlopers. Also, the determination of the physical extent of a cluster is problematic. Theoretically speaking, it is important to recall that the VME is not very sensitive to anisotropies  (e.g. Aceves \& Perea 1998) so its output can be more indicative of the equilibrium state of the system than of its true mass.  On other hand, the use of gravitational lensing methods, which are not hindered by the requirement of the cluster to be in equilibrium, can in conjunction with VME results assess the reliability of the latter. However mass determination by lensing methods are particularly affected if the cluster under investigation is elongated along the line-of-sight.

	A comment on the TC results for the VME is also pertinent here. Their projected mass estimator $V2$ is again only suited  when particles (e.g. galaxies) can be treated as test particles. If the latter is not true, the  application of $V2$ underestimates the actual mass. In fact, when such estimator was applied e.g. to our Hernquist $N$-body model we obtained $V2\approx 0.2$, a factor 5 lower than the true mass which is in agreement with the behaviour observed by TC.  We now turn to consider the continuous form of the virial theorem.

\subsection{Continuous Virial Theorem}

To derive the scalar virial theorem in its continuous form we multiply the Jeans hydrostatic equilibrium equation by $r\,$, and then integrate over the volume of the gravitational system (Binney \& Tremaine 1987). Using equation (2) for an isotropic velocity system, we have:
\begin{equation}
\int r \frac{{\rm d} \rho \sigma^2_r}{{\rm d}r}\, 4\pi r^2 \, {\rm d}r = - \int r \rho \frac{{\rm d} \varphi}{{\rm d}r}\, 4\pi r^2 \,{\rm d}r  \;.
\end{equation}
Integrating by parts the left hand side and expressing the right hand side in terms of the mass interior to $r$, we have:
\begin{equation}
r^3 \rho \sigma^2_r  - 3 \int^r r^2 \rho \sigma_r^2 \,{\rm d}r = 
- \int^r G r \rho M(r) \, {\rm d}r   \;.
\end{equation}
The {\it surface pressure} term is $3PV\equiv 4 \pi r^3 \rho \sigma_r^2\,$ (e.g. CYE). For future reference, we may estimate the position of the maximum of this term for the Dehnen's model, $3PV = G M^2 a r^3 (a+6r) (a+r)^{-6}/10$ : $\,r_{\rm max}=(7+\sqrt{65})a/8 \approx 0.65$, yielding a maximum value of $3PV_{\rm max} \approx 0.04\,$.

	It is readily verified that the Jeans equation holds at every radius, giving  e.g. for the Dehnen's model at each side of equation (9): $-3GM^2a r (r+a)^{-7}/4\pi$. Thus, neglecting the $3PV$ term, which comes from an integration by parts, would obviously produce an error in the mass estimate when  using Eq. (19) (e.g. The \& White 1986). Now, since the value of the mass $M(r)$ is inside an integral we cannot simply subtract the $3PV$ term point by point, or its maximum value, to an estimate of the mass using the VME (17) in order to obtain the correct value; but see below when this term is expressed in mass units (e.g. Girardi et al. 1998). This is because the $3PV$ term is local while the integrals in Eq. (19) and the mass are cumulative quantities.

	To quantify the effect of the $3PV$ term on the mass determination we should formally solve the integral equation (19). Fortunately, if one does not considers this surface pressure term the situation is simple. 
We have:
\begin{equation}
\int^r \left\{\,3r^2\rho\sigma_r^2 - G r \rho M(r)\,\right\}\,{\rm d}r = 0 \,,
\end{equation}
which we may solve readily, leading to an expression for the mass with no surface pressure term as follows:
\begin{equation}
M_{\rm NSP} (r) = \frac{3}{G} \, r \, \sigma_r^2 (r) \,.  
\end{equation}

\begin{figure}[!t]
\epsfysize=8cm
\epsfxsize=8.5cm
\epsffile{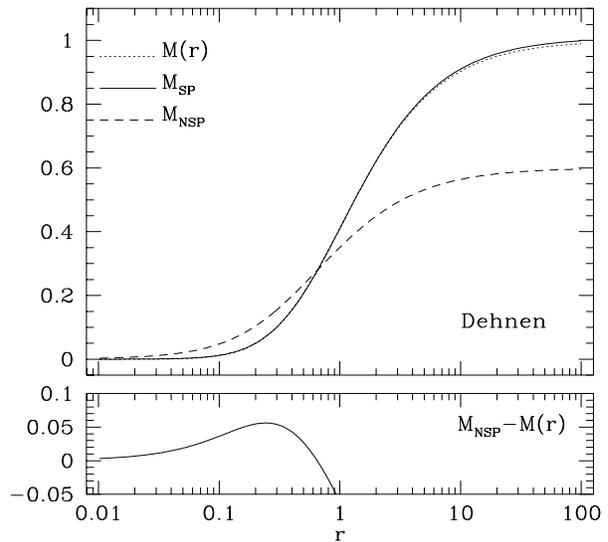}
\caption{Effect of neglecting the {\it surface pressure} ($3PV$) term in the continuous form of the virial theorem. $M_{\rm NSP}$ indicates that no $3PV$ term is considered in the solution of the integral equation (19). $M_{\rm SP}$ takes the referred term into consideration when actually solving the integral equation (19) for $M(r)$, and $M(r)$ is the theoretical Dehnen mass profile. Neglecting the $3PV$ term yields an overestimate of mass at  $r \lta R_{ \rm e}$ while an underestimate for $r \gta R_{\rm e}\,$. In the lower panel the difference $M_{\rm NSP} - M(r)$ is shown for a particular range of values.}
\end{figure}

	From Eq. (21) we can see that the neglect of the $3PV$ term in (19) does not always overestimates the mass, but can even underestimate it in realistic cases due to the radial dependence of the velocity dispersion. Indeed, for the extreme case of an isothermal sphere, where $\sigma^2_r$ is a constant, we have an overestimation since $M_{\rm NSP}(r) \propto r\,$. But, e.g. for the Dehnen's model where $\sigma^2_r \propto r^{-1}$ at large radii the mass will cease to increase settling to a constant value at some finite radius. 
	We may estimate the mass interior to $r=100$, using the expression for $\sigma_r^2$ in $\S2.1.1$, giving $M_{\rm NSP}(100) \approx 0.6\,$; i.e. an underestimation of the total mass by $\approx 40$\%.  Therefore, in general, an overestimation occurs only at the region of the system that may be considered nearly isothermal, e.g. $r \lta R_e$ (see Fig. 5), while an underestimation occurs outside of it. The maximum overestimate is approximately given by the numerical value of $3PV_{\rm max}$.

	Girardi et al. (1998) have introduced a correction term in Eq. (17) when partial sampling of a system is done based on an suggestion by CYE; namely, that an overestimate in mass occurs when applying the VME (eq. 17) due to the neglect of the $3PV$ term in equation (19). Girardi et al. have proposed the following formula to correct for virial mass estimates when partial sampling is done up to a boundary radius $b$
\begin{equation}
M_{\rm CV} = M_{\rm V} \left[ 1 - \frac{4 \pi b^3 \rho(b) \sigma_r^2 (b)}{ \int^b 4 \pi r^2 \rho \,{\rm d}r \,\times \, \sigma^2(<b)   }  \right] \,,
\end{equation}
where $\sigma(<b)$ refers to the integrated three-dimensional velocity dispersion, which we calculate in our numerical models as 
$$
\sigma^2 (<R) = \frac{3}{N(<R)} \sum_i^{N_<}  v_{z,i}^2    \, ,
$$
where $N(<R)$ refers to the total number of particles inside radius $R$ and the summation considers only the particles inside this $R$. 
The correction term in (22) expresses the $3PV$ term in `mass units' by dividing it by a term related to the total kinetic energy of the system up to the radius $b$.

	In Fig. 6 ({\it upper panel}) we show the computed projected mass profile using the VME by apertures including the $3PV$ correction term  ({\it dashed line}, $M_{\rm CV}$) as in equation (22) for our $N$-body equilibrium system. 
	The values of $\rho(r)$, $\sigma_r(r)$, and $M(r)$ were taken here directly from the theoretical model, hence representing an ideal observational situation, while the integrated velocity dispersion was calculated as indicated above. 	For comparison, the mass using the VME when the outer particles are considered, $M_{\rm O}$, is also shown. In the lower panel the corresponding differences are displayed.

\begin{figure}[!t]
\epsfysize=8.5cm
\epsfxsize=8.5cm
\epsffile{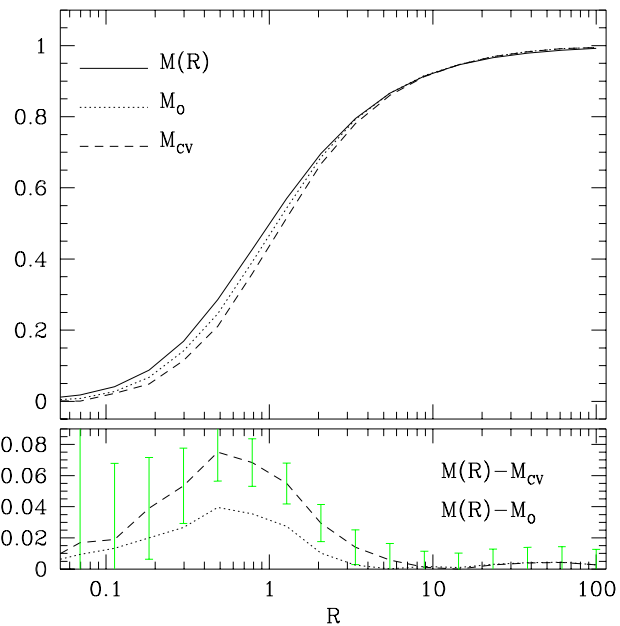}
\caption{Projected mass profile by the application of the VME by apertures, corrected by a surface pressure term for a Dehnen's model, $M_{\rm CV}$. $M(R)$ denotes the true projected mass for the model. In the lower panel, the differences between these quantities are shown. The difference $M(R)-M_{\rm CV}$ cannot be accounted by Poisson error bars.}
\end{figure} 

	From Fig. 6 it follows that both the masses obtained from the consistent application of the VME (eq. 17), and the correction due to the $3PV$ term introduced to aperture values of the VME (eq. 22) provide good estimates of the mass profile $M(R)$. 
	However the application of the VME to the $N$-body system yields a better result, within numerical errors, than Eq. (22). 
	We emphasize that in astronomical practice the VME does not require to assume any particular form of the radial velocity dispersion of galaxies, which is needed when constructing the $3PV$ correction in (22). This being perhaps the best feature of the virial mass estimator.

	Moreover, in an astronomical situation, in order to obtain the $3PV$ term one has to determine the number surface density of galaxies and then apply a deprojection integral to obtain the physical number density $n(r)$, or use the corresponding $\rho(r)$ of the fitted model.
 	Here, implicitly, some confidence is required that the equilibrium part of a cluster has been sufficiently sampled to obtain a reliable total virial mass $M$ and integrated velocity profile; as recognized by Girardi et al. (1998).  Hence, in this respect, the estimation of a mass profile using the
the correction term in (22) and that on the VME rest on the same working conditions, but the latter does not require any assumption on $\sigma_r(r)\,$ or its projected version. The PME correction term (6) is hindered by the same situation. 	
	Thus, we are lead to conclude that the VME is a reliable estimator of mass and mass profiles, having aside the virtue of being straightforward to apply.

	As the PME, the VME can overestimate the luminous matter of a cluster if its extent is small in comparison to the extent of the mass.  In this case one is led to conclude that the mass-to-light ratios of the visible parts of clusters may be importantly overestimated, say by $\sim 20$\%, if their outer parts are mainly dark.
	To conclude this section we recall that in deriving equation (17) the assumptions were the same as in the derivation of the PME, so these equations are valid in this context, as we have verified by numerical experiments that have correctly considered the potential energy term in Eq. 17.   In a future work we will consider the estimation of mass profiles of several clusters and compare the results with those obtained by other authors, both in the optical and X-ray region of the spectra.

\section{Conclusions}

We have investigated the behaviour of the PME and VME when applied to a  isotropic equilibrium system, and when partial sampling of this has been performed. 

We summarize our main conclusions as follows:

\begin{enumerate}

\item The projected mass estimator (PME) and the virial mass estimator (VME) yield accurate results for the total mass of a gravitational system, provided that the whole system is sampled, is in equilibrium and has an isotropic velocity dispersion. Moreover, the VME can provide an accurate mass profile under the previous conditions without requiring the isotropy condition.

\item The PME overestimates the mass if the sampled region is a small portion of the total system. The maximum error occurs around the total system's effective radius $R_{\rm e}$, and depends on the density profile. For realistic profiles this overestimate can be up to $\approx 20$\%. 
 	The VME yield similar errors  when partial sampling of a complete system is made. This is because the summation $\sum 1/R_{ij}$ in the VME (eq. 17) does not includes all members of the system, but only those we can \emph{observe}.

\item  A recently introduced correction term based on the surface pressure term of the continuous virial theorem is found to perform, for practical matters, as well as the VME under the same kind of hypotheses on the stellar system. In applications, e.g. to compute the mass distribution of clusters using galaxy data, however, the VME should be preferred here since it does not require any assumptions on the radial velocity dispersion of galaxies.
	Discrepancies found in the use of the VME, or the PME, in some $N$-body simulations are attributed to treating particles as test-particles and/or to anisotropies present in the numerical clusters.

\end{enumerate}

\begin{acknowledgements}

H.A. is grateful to the Spanish Ministry of Foreign Affairs for a MUTIS Studentship. J.P. acknowledges financial support from the Spanish DGICYT project PB96-0921.  Dr. Joshua Barnes is thanked for making available his {\sf TreeCode}, and L. Verdes-Montenegro and A. Del Olmo for their comments which helped to improve the presentation of the paper. The Editor and an anonymous referee are also thanked for helpful comments. 

\end{acknowledgements}

\end{document}